\documentclass[letterpaper, 10 pt, journal, twoside]{ieeetran}

\usepackage{graphicx} 
\usepackage{framed}
\usepackage{multirow}
\usepackage{booktabs}
\usepackage{url}
\usepackage{fancyvrb}
\usepackage[T1]{fontenc}
\usepackage{tabularx}

\begin{document}

\title{Learning from Success and Failure:\\Acquiring Adaptive Dialogue Strategies for Social Robots}

\author{Sanae Yamashita, Yuki Okafuji
    \thanks{Manuscript received: November 26, 2025; Revised February 21, 2026; Accepted: May 26, 2026.}
    \thanks{This paper was recommended for publication by Editor Tetsuya Ogata upon evaluation of the Associate Editor and Reviewers' comments.}
    \thanks{Sanae Yamashita and Yuki Okafuji are with CyberAgent, Tokyo, Japan and The University of Osaka, Osaka, Japan {\tt\footnotesize yamashita\_sanae\_xa@cyberagent.co.jp} {\tt\footnotesize okafuji\_yuki\_xd@cyberagent.co.jp}}
    \thanks{Digital Object Identifier (DOI): 10.1109/LRA.2026.3704002}
}

\markboth{IEEE Robotics and Automation Letters. Preprint Version. Accepted June, 2026}
{Yamashita \MakeLowercase{\textit{et al.}}: Acquiring Adaptive Dialogue Strategies for Social Robots}

\maketitle

\begin{abstract}
Traditional dialogue systems for social robots require both dialogue strategies and user attribute recognition, each demanding specialized expertise. However, data collection is costly in real-world deployments, and the resulting datasets often include many failure cases. In this study, we aim to automate the acquisition of dialogue strategies by leveraging both successful and failed interactions using a vision-language model (VLM) and a large language model (LLM). We propose an architecture in which user attributes, recognized by the VLM, along with dialogue history, are fed into the LLM to generate dialogue strategies tailored to specific user attributes. We extracted dialogue strategies from an interaction dataset collected through a field experiment and evaluated their effectiveness. The results demonstrate that explicitly representing failure strategies complements success strategies and improves performance. Our findings highlight a practical pipeline for constructing and maintaining an interpretable strategy repository from in-the-wild deployment logs by recycling abundant failure interactions as reusable constraints, ultimately reducing the development cost of social robots.
\end{abstract}


\section{Introduction}
\IEEEPARstart{S}{ocial} robots have attracted increasing attention in customer service applications, such as route guidance and product recommendations~\cite{okafuji2022behavioral}. These robots interact directly with users, rendering them promising for real-world deployment. Real-world deployments of social robots inevitably involve both successful and failed interactions. Recent work has used large language models (LLMs) to acquire dialogue strategies from successful human-human conversations~\cite{xie2024few}. However, such approaches typically assume access to many successful cases, which is unrealistic for social robots: in customer service, for example, customers interact differently with human staff and robots~\cite{kanda2010communication}, and success strategies for social robots must be established through in-the-wild deployments. As a result, collected datasets inevitably contain a substantial number of failures as well as successes.

Failures themselves can provide informative signals for improving interaction. In goal-oriented dialogue systems, learning from failure experiences has been shown to enhance the robustness of dialogue policies~\cite{lu2019learnfailure}, and field studies of long-term human-robot interaction report that failures and their recoveries critically shape user perceptions of social robots~\cite{delduchetto2023fail}. These findings suggest that, if failure cases can be exploited appropriately, social robots may acquire effective dialogue strategies more efficiently than approaches that rely only on success cases. Nevertheless, how to systematically utilize failure cases for learning dialogue strategies for social robots remains underexplored. Importantly, our objective is to convert failure episodes into explicit \emph{[Failure]} if--then strategies that describe interaction anti-patterns (what the robot should avoid) under a given user context. Because these strategies are designer-readable and editable, they can be inspected, revised, and reused after deployment; moreover, they are provided as negative constraints during LLM-based utterance generation to prevent repeated failure behaviors.

At the same time, social robots often adapt their behavior based on user attributes and states to perform tasks effectively~\cite{kass1988modeling}. Traditionally, they have been implemented using a pipeline that recognizes user attributes~\cite{FONG2003143}, selects a strategy based on these attributes~\cite{janarthanam2014adaptive}, and executes behaviors accordingly. Attribute recognition typically relies on multiple classification models for age, gender, emotions, and interests~\cite{VINCIARELLI20091743}, which are costly to develop and maintain. Dialogue strategies in such pipelines are often rule-based owing to the complexity of modeling dialogue interactions~\cite{okafuji2022behavioral}, resulting in extensive manual rule design and limited scalability.

This study aims to automatically acquire dialogue strategies tailored to user attributes by leveraging both successful and failed interactions. We propose an architecture that integrates a vision-language model (VLM) to recognize user attributes and an LLM to generate dialogue strategies based on user attributes and dialogue history, incorporating both successful and failed interactions. Note that this study does not aim to acquire a single optimal dialogue strategy, but rather to accumulate reusable strategies derived from diverse interaction experiences. In our experiment, dialogue strategies were extracted from an interaction dataset collected in a field experiment in which a robot performed a route guidance task. The acquired strategies were then evaluated to assess their effectiveness.

Compared to prior VLM or LLM-based conversational robot frameworks, our technical novelty lies in (i) explicitly converting both success and failure interactions into designer-editable natural-language if--then strategies, (ii) learning them entirely offline from in-the-wild deployment logs without online reinforcement learning, and (iii) stabilizing long-term strategy accumulation through clustering-based aggregation.

The primary contributions of this study are as follows:
\begin{itemize}
\item We propose a VLM/LLM architecture that acquires and organizes a designer-readable and editable dialogue-strategies (natural-language if--then rules) from in-the-wild interaction, and we introduce an offline evaluation method to assess the resulting strategies.
\item Additionally, we demonstrate, through a field experiment on robot-assisted route guidance, that leveraging both successful and failed cases leads to the acquisition of more effective dialogue strategies.
\item Furthermore, we analyze the role of explicit user attributes in this setting, showing that their impact on prediction performance is limited and discussing implications for the ethical use of attribute-based adaptation in social robots.
\end{itemize}

\section{Related Work}
\label{sec:related_work}

\subsection{Automatic Acquisition of Behavioral Strategies}
In robotics and dialogue systems, research on automatically acquiring behavioral strategies with LLMs is rapidly advancing. Several studies in robotic control and task planning map natural language instructions to executable actions, enabling robots to respond to novel commands~\cite{yao2023react}. Liang et al.~\cite{liang2023code} proposed an approach in which LLMs generate strategies as source code that is directly applied to robot control. Agent-based exploration studies leverage LLM-derived knowledge for dynamic behavior selection~\cite{wang2023voyager}. LLMs have also been applied to dialogue strategy acquisition. Xu et al.~\cite{xu2024language} developed a Werewolf dialogue system in which LLMs generate possible next actions, from which the system derives its strategy. Xie et al.~\cite{xie2024few} proposed a method that uses expert-curated ideal dialogue data to acquire successful dialogue strategies with an LLMs.

However, real-world social-robot deployments yield many failures, and how to exploit failure cases for dialogue strategy acquisition remains underexplored. Our method does not rely on expert-curated or predominantly successful dialogues; instead, it leverages both successful and failed interactions, reducing curation cost while expanding training signals. 

One approach to learning dialogue strategies from both successful and failed interactions is reinforcement learning (RL). RL can learn dialogue strategies from success/failure rewards~\cite{su2016continuously,lu2019learnfailure}, but typically requires online trial-and-error and careful reward design, which is difficult in in-the-wild deployments. In contrast, our architecture learns strategies entirely offline from interaction logs, avoiding online exploration and reward engineering.

\subsection{User Attribute Recognition}
For effective user-adaptive dialogue strategies, a broad range of user attributes has to be considered. Previous studies have explored attributes including age, gender, educational background~\cite{mctear2002spoken}, emotions~\cite{paiva2017empathy}, facial expressions~\cite{devault2014simsensei}, personality traits~\cite{yamamoto2023character}, knowledge level, urgency~\cite{komatani2003flexible}, and engagement~\cite{del2022learning}. Adapting dialogue strategies to personality traits can improve user experience and facilitate smoother interactions~\cite{yamamoto2023character}. These findings motivate dialogue systems that adapt to user attributes in service-robot tasks.

Recent research has explored VLMs for user attribute recognition. Flamingo~\cite{alayrac2022flamingo}, a large-scale model that processes both text and visual inputs, shows potential for comprehensively understanding human and environmental contexts. Studies have also investigated generating natural language descriptions from detected individuals and objects in images~\cite{li2023blip2}, suggesting that VLM-generated attributes can enhance real-time dialogue responses. However, the extent to which automatically recognized attributes contribute to dialogue strategy acquisition in real-world social-robot tasks remains unclear.

\begin{figure*}[!t]
\centering
\includegraphics[width=\linewidth]{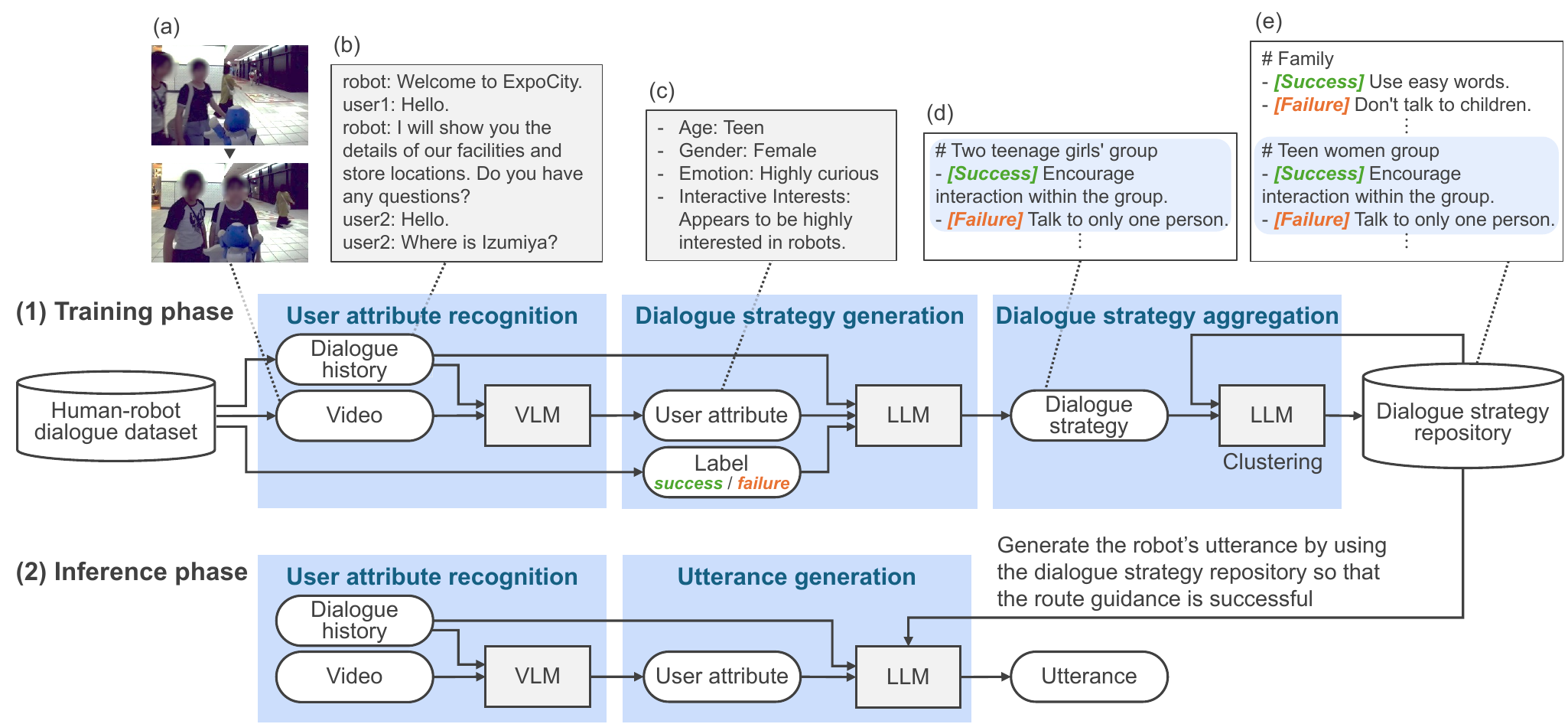}
\caption{Proposed dialogue strategy architecture that acquires both successful and failed dialogue strategies based on user attributes. (1) Training phase comprises three key steps: user attribute recognition, dialogue strategy generation, and dialogue strategy aggregation. (a) Videos are sourced from a human-robot interaction dataset. (b) The system utilizes dialogue history as input. (c) User attributes are recognized by processing both the dialogue history and video through the VLM. (d) Dialogue strategies are generated by inputting the identified user attributes and corresponding labels into the LLM. (e) The newly generated dialogue strategies are combined with the accumulated dialogue strategy repository through the LLM, resulting in an updated dialogue strategy repository. (2) In the inference phase, the LLM recognizes user attributes in the same manner, receives the user attributes and the dialogue strategy repository, and generates the robot’s utterance to make the dialogue successful, taking into account successful and failed strategies.}
\label{fig:approach}
\end{figure*}

\section{Architecture}
\label{sec:approach}
This section presents an architecture designed to automatically acquire both successful and failed dialogue strategies based on user attributes. Fig.~\ref{fig:approach} provides an overview of the system, which processes dialogue history, visual data, and success/failure labels to derive and iteratively refine strategies in a dialogue strategy repository. By continuously updating the repository across interactions, the system accumulates a comprehensive set of strategies. The architecture comprises three key components: user attribute recognition, dialogue strategy generation, and dialogue strategy aggregation. The following sections explain each component in detail.

\subsection{User Attribute Recognition}
To enable unified inference of multiple user attributes within a single model, we adopt a VLM-based recognition approach. The VLM, implemented using GPT-4o, processes dialogue history and visual data to recognize user attributes via zero-shot inference and outputs them as text (Fig.~\ref{fig:approach}~(c)). GPT-4o does not support direct video input; hence, video data are converted into one-second interval images, encoded in base64, and input sequentially.

Following prior research, this method recognizes key user attributes, including age, gender, emotions, and interest in the robot~\cite{Foggia2024SoftBiometrics}. Although nonverbal behaviors such as gestures, body orientation, and interpersonal distance may also be useful, this method treats them as cues for measuring interest and implicitly integrates them into interest in the robot. For group interactions, additional attributes such as group size, relationships among members, overall group interest level, and interaction dynamics are also identified~\cite{sakaguchi2022estimation}.

The prompt first distinguishes whether the interaction partner is an individual or a group. For an individual, only individual attributes are output; for a group, group-level attributes are additionally extracted. To enhance dialogue-strategy reusability, the prompt extracts general attributes by excluding situation-dependent information such as location. The output is standardized as a bullet-point list.

\subsection{Dialogue Strategy Generation}
After recognizing user attributes, the LLM (GPT-4o) generates successful and failed dialogue strategies in text form by processing user attributes, dialogue history, and success/failure labels via zero-shot inference. The concise prompt used for dialogue strategy generation is shown in Table~\ref{tab:prompt_template}. Following a Chain-of-Thought-style structure, the prompt first instructs the model to organize user attributes, then to enumerate robot behaviors that are considered to have provided high user satisfaction, and finally to describe what kinds of robot behaviors increase satisfaction for users or groups with particular attributes. Depending on the experimental condition, the model is instructed to output only successful strategies, only failed strategies, or both. The output format is unified as bullet points.

The generated dialogue strategies follow an if--then structure, as illustrated in Fig.~\ref{fig:approach}~(d). Each user attribute is associated with multiple dialogue strategies. To distinguish between success and failure cases, we label the strategies with the prefixes \texttt{[Success]} and \texttt{[Failure]}.

In the inference phase, the strategy repository is provided to the LLM with an explicit instruction to follow success strategies while treating failure strategies as explicit negative constraints on behaviors to avoid.

\begin{table}[t]
    \caption{Prompt Template (Condensed)}
    \label{tab:prompt_template}
    \centering
    \small
    \setlength{\tabcolsep}{4pt}
    \renewcommand{\arraystretch}{1.1}
    \begin{tabularx}{\columnwidth}{@{}lX@{}}
    \hline
    \textbf{Field} & \textbf{Content} \\
    \hline
    Input &
    \#User attributes (cluster label + bullet attributes), \#Dialogue history (robot/user turns) , \#Outcome label (Success/Failure). \\
    Instruction &
    (1) Concisely organize the user attributes. (2) Enumerate robot behaviors in the dialogue history that are considered to have provided high user satisfaction. (3) Describe what kinds of robot behaviors increase satisfaction for users/groups with particular attributes. \\
    Output format &
    Unified as bullet points. Example: \newline
    \texttt{\# <organized attributes>} \newline
    \texttt{- [Success] <behavior strategy>} \newline
    \texttt{- [Failure] <behavior strategy>} \\
    \hline
    \end{tabularx}
\end{table}

\subsection{Dialogue Strategy Aggregation}
After being generated, dialogue strategies are stored and continuously updated in the dialogue strategy repository (Fig.~\ref{fig:approach}~(e)). Here, we identify two issues in managing the dialogue strategy repository. First, because dialogue strategies are generated for each dialogue, naively appending them causes the repository to grow as training data increases, resulting in a long prompt when an LLM generates utterances in a subsequent phase. Second, since strategies are generated independently for each dialogue, similar strategies are repeatedly produced, leading to redundancy in the repository. In a preliminary attempt to address this redundancy, we applied LLM-based aggregation with overwriting; however, this approach proved unstable, causing unintended loss of strategies. To mitigate this, we adopt a clustering-based approach using text embeddings to systematically integrate newly acquired strategies with existing ones.

This process comprises three steps: generating text embeddings, clustering, and naming clusters. First, user attributes and dialogue strategies are converted into embeddings using Japanese Simple-SimCSE\footnote{\url{https://github.com/hppRC/simple-simcse-ja}}. Next, DBSCAN is applied to these embeddings to group similar attributes and strategies into clusters. Finally, names are assigned to these clusters by feeding the grouped user attributes into an LLM (GPT-4o), which generates representative attribute labels via zero-shot inference. The model is instructed to summarize the common user attributes within each cluster into a short representative label. This structured aggregation process ensures a more stable and efficient organization of dialogue strategies while minimizing redundancy. This aggregation process is performed each time new dialogue strategies are generated.

\section{Experiment}
\label{sec:experiment}

\begin{figure}[t]
    \centering
    \includegraphics[width=0.87\linewidth]{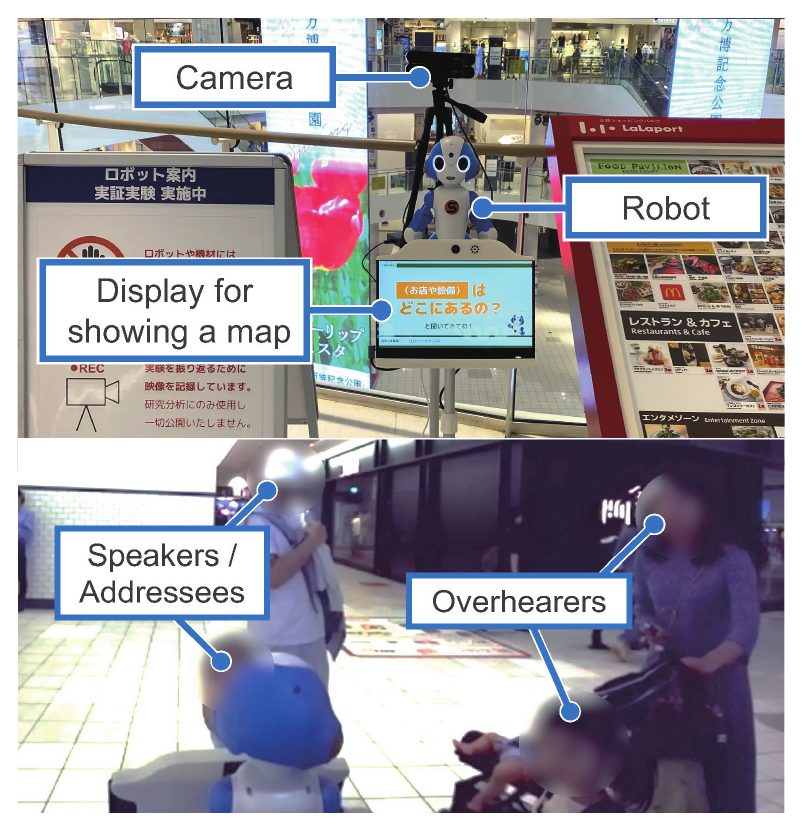}
    \caption{Sota robot, designed for route guidance and an example of a video. The display explicitly provides directional assistance, ensuring users are aware of its functionality. In the recorded video, two users interact with the robot, while another two users watch as overhearers.}
    \label{fig:experiment}
\end{figure}

\subsection{Dataset Collection in Field Experiment}
\label{sec:dataset_collenction}
We collected a dialogue dataset for a route guidance task through a human-robot field experiment, approved by the ethical committee of The University of Osaka. Data collection took place over two days at a large shopping mall, where the Sota robot (Fig.~\ref{fig:experiment}) interacted with visitors. The robot provided route guidance, engaged in small talk, and covered a range of dialogue topics. It used GPT-4o to answer user questions with route explanations, with its prompt including only store-related information and no predefined conversational strategies. 

Most dialogues lasted approximately one minute, with an average of four utterances per dialogue. An example dialogue is shown in Fig.~\ref{fig:approach}(b). Due to the nature of the route guidance task, dialogues were generally short, and the topic rarely changed during the interaction. The dataset includes transcribed dialogue histories at the utterance level and videos capturing both the user and the robot during interactions (Fig.~\ref{fig:approach}~(a)). Each of the 236 recorded dialogues was labeled as either a success or failure based on whether route guidance was successfully provided. A successful interaction was defined as one where the robot offered route guidance in response to a user's question, while a failed interaction occurred when the user either did not ask a question or did ask one but did not receive guidance. To streamline evaluation, we adopted a simple success criterion route guidance provision to facilitate the evaluation of acquired dialogue strategies. However, this definition could be expanded to include more complex metrics, such as user satisfaction. To assess labeling reliability, two annotators independently labeled the dialogues using the criterion above; inter-annotator agreement was high (Cohen's $\kappa$ = 0.9).

Of the 236 collected dialogues, 196 were used for training and 40 for testing. The training set comprised 88 successful and 108 failed cases, while the test set included 20 successful and 20 failed dialogues.

\subsection{Experimental Setup}
Dialogue strategies are derived from the collected experimental data using the proposed architecture. A total of 9 experimental conditions are established by varying two factors: user attribute recognition methods (three types) and the type of acquired dialogue strategy (three types).

For user attribute recognition, three approaches are considered. The first is VLM-based recognition, which identifies user attributes from both dialogue history and video. The second is LLM-based recognition, which extracts attributes solely from dialogue history. The third is a no-recognition (None) condition, where user attributes are not identified; in this case, attribute-based aggregation is not performed, and all dialogue strategies are grouped into a single cluster.

The experiment explores three types of dialogue strategies. In the both-strategy condition, both success and failure strategies are extracted from all dialogues. The success-only condition extracts only successful strategies, while the failure-only condition focuses exclusively on failure strategies.

\begin{table}[t]
    \scriptsize
    \caption{Examples of Short and Long Inputs Used in Evaluation}
    \label{tab:input}
    \centering
    \begin{tabular}{p{8cm}}
        \toprule
        \textbf{Short Input} \\
        \midrule
        robot: Welcome to EXPOCITY! I can help you find stores and facilities and provide details about their locations. Do you have any questions? \\
        user: What is your name? \\
        \midrule
        \textbf{Long Input (added)} \\
        \midrule
        robot: My name is Sota. Nice to meet you. Feel free to ask me about the facility. \\
        user: I want to eat ramen. \\
        robot: I provide information about the locations of facilities and stores. Please feel free to ask. \\
        user: Where is Shisokutenka? \\
        \bottomrule
    \end{tabular}
\end{table}

\subsection{Evaluation Method}
We evaluate both the quality of the acquired dialogue strategies and the quality of the robot’s next utterances generated using those dialogue strategies.

\subsubsection{Offline Prediction Task}
One way to evaluate dialogue strategies is by generating responses based on acquired strategies and assessing them using metrics such as task success rate and user satisfaction~\cite{walker2000application}, as well as turn count~\cite{scheffler2022automatic}. Additionally, other dialogue system evaluation metrics, such as consistency and informativeness~\cite{finch-choi-2020-towards}, can be applied. However, these methods rely on online interactions; there is no established approach for evaluating dialogue strategies offline.

To address this limitation, this study proposes an offline evaluation method for dialogue strategies. To assess the effectiveness of the proposed architecture, we introduce a prediction task in which the acquired dialogue strategy repository is used to predict whether a given abbreviated test dialogue will ultimately succeed or fail. This approach assumes that a well-refined dialogue strategy repository should enable accurate prediction of dialogue outcomes based on ongoing dialogue history. For performance evaluation, the LLM is provided with user attributes, dialogue history, and the final dialogue strategy repository obtained under each experimental condition. The model then outputs a binary prediction: Success or Failure.

To analyze the effect of input length on prediction accuracy, we divide the test dialogue histories into short and long input conditions (Table~\ref{tab:input}). Short inputs include dialogue history up to the user's first utterance (ranging from 1 to 4 turns), while long inputs extend up to the user's first question (ranging from 3 to 6 turns). This setup enables comparison between early- and later-stage conversation-based predictions. The dataset comprises 20 short-input dialogues and 20 long-input dialogues, with 10 successful and 10 failed cases in each category. For reference, six human evaluators performed the same prediction task using the identical short/long video clips and dialogue histories under the same success definition; predictions were made independently in randomized, condition-blind settings, and we report the average F1.

\subsubsection{Utterance Generation Task}
Separately from the prediction task, we conduct a human preference study on next utterances generated from the acquired dialogue strategies. As shown in the inference phase in Fig.~\ref{fig:approach}, user attributes were first recognized from the dialogue history and video. Next, the dialogue history, recognized user attributes, and the acquired dialogue strategy repository were provided as input, and the model was prompted to generate the robot’s next utterance so that the route guidance would be successful. The dialogue history and video were taken from the middle of the interaction and truncated just before the robot’s utterance. The dialogue strategies used were those acquired under the VLM Both, LLM Both, and None Both conditions.

For evaluation, six participants watched eight representative video clips together with three candidate next utterances (anonymized and presented in random order) and selected the most appropriate one, defined as the utterance most likely to lead to successful route guidance.

\begin{table}[t]
\scriptsize
\caption{Examples of Dialogue Strategies}
\label{tab:dialogue_strategy_example}
\centering
\begin{tabular}{p{8cm}}
\toprule
\# Parents and young children\\
- [Success] Offer simple games or fun information\\
- [Success] Use a child-friendly tone\\
- [Success] Give parents ideas for activities that children will enjoy\\
- [Failure] The robot does not adapt to the user's individual interests and stubbornly sticks to its predefined role\\
- [Failure] Ignores the child's interests and rigidly adheres to the dialogue\\
- [Failure] Fails to consider the parent's perspective and refuses the child's requests to play\\
\bottomrule
\end{tabular}
\end{table}

\begin{table}[t]
    \caption{Average Number of Clusters and Dialogue Strategies}
    \label{tab:count_of_user_attributes_and_strategies}
    \centering
    \begin{tabular}{lcc}
    \toprule
    ~ &  & \textbf{Strategies}\\
    ~ & \textbf{Clusters} & \textbf{Success / Failure / Total}\\
    \midrule
    VLM Both & 90.8 & 429.6 / 471.3 / 901.0\\
    VLM Success & 55.6 & 428.5 / 0.0 / 428.5\\
    VLM Failure & 55.3 & 0.0 / 451.6 / 451.6\\
    \midrule
    LLM Both & 82.8 & 408.0 / 322.6 / 730.6\\
    LLM Success & 54.1 & 404.6 / 0.0 / 404.6\\
    LLM Failure & 46.5 & 0.0 / 327.6 / 327.6\\
    \midrule
    None Both & 1 & 286.0 / 227.1 / 513.1\\
    None Success & 1 & 285.8 / 0.0 / 285.8\\
    None Failure & 1 & 0.0 / 236.3 / 236.3\\
    \bottomrule
    \end{tabular}
\end{table}

\begin{figure}[t]
\centering
\includegraphics[width=\linewidth]{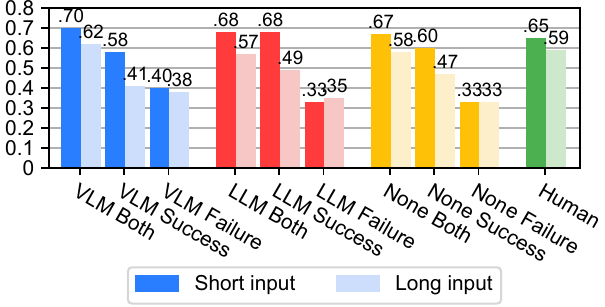}
\caption{Macro F1 score for the evaluation results of success/failure prediction.}
\label{fig:result_f1score}
\end{figure}

\subsection{Results in Offline Prediction Task}
\label{sec:results_in_offline_prediction_task}
For each of the 9 experimental conditions, training was conducted six times using the same training dataset; the acquired dialogue strategies were evaluated using the test data. Both the VLM and LLM utilized OpenAI's \texttt{gpt-4o-2024-05-13}. As a result, for example, from a dialogue with a parent and a child, a dialogue strategy as shown in  Table~\ref{tab:dialogue_strategy_example} was obtained.

Table~\ref{tab:count_of_user_attributes_and_strategies} presents the average number of clusters and dialogue strategies acquired. When using either VLM or LLM, approximately 50 clusters were formed, each containing corresponding success and failure strategies. Within the same condition, the number of acquired success and failure strategies remained relatively balanced. Comparing user attribute recognition methods, LLM yielded fewer strategies than VLM, while the None condition resulted in the fewest.

Fig.~\ref{fig:result_f1score} illustrates the evaluation results of success/failure prediction in terms of F1-score. Regardless of the user attribute recognition method (VLM, LLM, None), results across attribution types showed similar trends. The most notable finding was that acquiring both success and failure strategies led to higher F1-scores compared to conditions where only one type of strategy was acquired. In particular, the VLM Both condition for short-input achieved the highest score, suggesting that incorporating failure strategies provides complementary negative guidance and improves the strategy repository beyond relying on success strategies alone. Conversely, conditions that acquired only failure strategies consistently yielded lower scores.

As a supplementary comparison, human raters also achieved scores as high as the Both condition, but slightly lower than the highest-scoring VLM Both setting.

\subsection{Results in Utterance Generation Task}
To complement the quantitative prediction results, we qualitatively analyzed utterances generated from the learned dialogue strategies. We focused on the most frequent interaction pattern where a parent and young child approached the robot, and the parent encouraged the child to talk to it.

First, we compared the three strategy types (Both, Success, Failure). In the Both condition, the robot produced an utterance such as: \textit{``Shall I explain the shops and activities in this facility? I can tell you about things that you and your child can enjoy together!''} This explicitly refers to both the parent and the child and aims to draw out the child’s interest, reflecting success strategies such as showing empathy and failure strategies such as lacking elements that capture children’s interest. Concretely, for the parent--child cluster, the repository contains success-side guidance to propose activities that children would enjoy, while the corresponding failure-side entries highlight anti-patterns such as ignoring the child's interests (Table~\ref{tab:dialogue_strategy_example}). The Both-condition utterance integrates these by explicitly addressing success and failure strategies. In contrast, the Success and Failure conditions yielded more straightforward suggestions to the parent, e.g.: (Success) \textit{``Are you curious about what kinds of shops and facilities are nearby? We can check the map together.''} (Failure) \textit{``Hello! Shall I introduce some interesting facilities? There are activities children can enjoy.''} These utterances mainly propose route guidance and do not clearly treat the child as an active conversational partner.

We also compared user attribute recognition methods (VLM Both, LLM Both, None Both). In the VLM Both condition, the system generated the previously mentioned utterance named Both, explicitly mentioning the parent-child relationship and proposing activities. In contrast, the LLM Both and None Both conditions produced more generic questions such as: (LLM) \textit{``Is there any specific place or shop you are looking for? Please feel free to tell me.''} (None) \textit{``Is something the matter? If there is anything you need help with, please let me know.''} These do not explicitly refer to the family relationship. Overall, even though user attributes did not substantially improve prediction accuracy, they still influenced how the robot phrased its utterances and how specifically it adapted to the parent-child scenario.

The distribution of participant selections was as follows: VLM Both received 19 votes (39.6\%), LLM Both received 17 votes (35.4\%), and None Both received 12 votes (25.0\%). These results indicate that VLM Both and LLM Both were able to generate more appropriate utterances than None Both.

\section{Analysis and Discussion}
\label{sec:analysis}
The previous section demonstrated that incorporating failure cases enhances the acquisition of effective dialogue strategies. However, within the scope of this experiment, no clear effect of user attributes was observed. Therefore, this section examines the accuracy of user attribute recognition, evaluates the necessity of it, and explores the impact of input length on prediction accuracy. We also conduct an evaluation under a different robot configuration.

\subsection{Recognition Accuracy of User Attribute}
To assess the accuracy of user attribute recognition by the VLM and LLM, all authors reviewed the 40 test dialogues, comparing the model-recognized attributes with video recordings. In cases where the authors' interpretations differed, they discussed and established a consensus on the correct answer criteria. The recognition results were classified into three categories: correct recognition, incorrect recognition, and unrecognized attributes.

Table~\ref{tab:vlm_llm_user_attributes_accuracy} presents the accuracy of user attribute recognition using the VLM. The results show that, except for gender, attributes were identified with over 60\% accuracy, highlighting the VLM’s effectiveness in capturing nonverbal cues. For instance, attributes such as ``multiple children approaching the robot'' or ``high overall group interest,'' which were challenging to infer solely from dialogue history, were accurately recognized from video data. However, the highest rate of incorrect recognition was observed for group size, where the model frequently misidentified overhearers as part of the interacting group. For example, the VLM recognized four users participating in the dialogue from the video in Fig.~\ref{fig:experiment}, but in reality only two participated and the other two are overhearers.

The LLM relies exclusively on dialogue history; hence, it often failed to determine  whether the user was part of a group, leading to a high number of cases where group information was unrecognized. Similarly, attributes such as age and gender were difficult to infer without explicit verbal cues, resulting in frequent non-recognition. However, for attributes related to emotions and interest in the robot, the LLM achieved approximately 50\% accuracy, suggesting that certain verbal cues within the dialogue history provided a basis for reasonable inference.

\begin{table}[t]
\centering
\caption{Recognition Accuracy of User Attribute by VLM and LLM. Each Value Represents Correct / Incorrect / Unrecognized.}
\label{tab:vlm_llm_user_attributes_accuracy}
\begin{tabular}{lcc}
\toprule
~ & \textbf{VLM} & \textbf{LLM}\\
\midrule
Age & 0.60 / 0.00 / 0.40 & 0.10 / 0.01 / 0.89 \\ 
Gender & 0.46 / 0.02 / 0.51 & 0.01 / 0.00 / 0.99 \\ 
Emotions & 0.62 / 0.05 / 0.33 & 0.40 / 0.08 / 0.51 \\ 
Interest level & 0.63 / 0.06 / 0.31 & 0.50 / 0.04 / 0.46 \\  
Group size & 0.55 / 0.32 / 0.13 & 0.13 / 0.10 / 0.77 \\ 
Group relationship & 0.74 / 0.03 / 0.23 & 0.13 / 0.06 / 0.81 \\ 
Group interest level & 0.71 / 0.13 / 0.16 & 0.16 / 0.06 / 0.77 \\ 
Interaction among members & 0.65 / 0.16 / 0.19 & 0.16 / 0.03 / 0.81 \\ 
\bottomrule
\end{tabular}
\end{table}

\subsection{Necessity of User Attribute Recognition via VLM}
One of the key objectives of this study was to acquire dialogue strategies that account for user attributes. To reassess the necessity of user attribute recognition, it is essential to examine how user behavior affects dialogue strategies. Previous research suggests that user behavior is a critical factor in interaction outcomes~\cite{kass1988modeling}. To explore this further, we collected comments from human evaluators with high prediction accuracy, highlighting the behavioral cues they relied on.

Human evaluators first noted the user's movements early in the interaction. For instance, when a user observed the robot from a distance before gradually approaching, it was interpreted as a sign of positive interest or familiarity with the robot's behavior, leading evaluators to predict a higher likelihood of a successful interaction. Such nonverbal behaviors are difficult to capture from textual dialogue history but can serve as valuable cues when recognized from video using a VLM. Conversely, behaviors such as turning away from the robot's P-space~\cite{kendon1990conducting} or a group member displaying boredom by using a smartphone were frequently identified as indicators of likely interaction failure. Prior research also suggests that the behavior of one user can influence others within the group~\cite{sakaguchi2022estimation}. As these are nonverbal cues, they are difficult to infer without analyzing video through a VLM.

These findings indicate that incorporating nonverbal information enables a more detailed understanding of engagement, interests and group interaction dynamics as user attributes and states, potentially enhancing the accuracy of dialogue success or failure prediction and improving the selection of appropriate dialogue strategies. However, in the test data, the VLM observed nonverbal behavior in only seven conversations (less than 20\% of the total), and its recognition of user behaviors that affect interaction success or failure was insufficient. This limited presence of nonverbal cues likely contributed to the minimal difference in task success prediction performance across conditions.

Furthermore, while the VLM outperformed the LLM in recognizing user attributes such as age, gender, and group interactions, the lack of a significant difference in task success prediction performance among the VLM, LLM, and None conditions suggests that the simplicity of the task may have diminished the impact of user attributes. In such a straightforward task, even with diverse user attributes, behavioral variations were limited, reducing their influence on task success prediction performance.

Beyond predictive performance, these results also raise questions about which user attributes should be modeled. In our architecture, the VLM recognizes a broad set of attributes, including identity-related characteristics such as age and gender, as well as more transient states such as emotions, individual interest in the robot, and group-level engagement. Identity characteristics of this kind are typically considered sensitive, particularly when they are inferred automatically from visual data. Prior work in human-robot interaction has shown that representing, recognizing, and reasoning over identity characteristics (e.g., age, gender, race) can introduce ethical risks such as misclassification, reinforcement of social stereotypes, and discriminatory treatment of users~\cite{williams2023eye}. These concerns are especially salient in public-space deployments with diverse users, as in our setting. Note that the time required for user attribute recognition using the VLM, which was conducted offline was approximately 39s ($\pm$12) per dialogue, mainly because the entire interaction was input to the VLM. Although this latency may not be as large in practical deployments, faster inference and mechanisms to maintain user engagement during processing would still be necessary. This also serves as a reason to reconsider the necessity of user attribute recognition.

For this relatively simple route guidance task, our findings indicate that the practical benefit of explicitly modeling identity characteristics is limited: task success prediction did not substantially improve when attributes such as age and gender were provided, and effective dialogue strategies could still be acquired by relying on task context and observable user behavior. At the same time, the VLM occasionally misrecognized identity attributes, which could lead the robot to condition its behavior on incorrect or overly essentialized representations of users. Therefore, our results support a design direction in which robots prioritize task-relevant, transient cues (e.g., engagement, interest, group interaction dynamics) over persistent identity categories. Focusing on behavioral and interactional signals may still enable meaningful adaptation of dialogue strategies while avoiding the need for explicit modeling of sensitive identity attributes; from this perspective, our study provides empirical support for minimizing reliance on explicit identity characteristics in everyday service-robot deployments.

\subsection{Effect of Input Length on Prediction Accuracy}
When predicting dialogue success or failure using short and long input conditions, short inputs consistently yielded higher prediction accuracy (Fig.~\ref{fig:result_f1score}). Despite containing more information, long inputs resulted in lower performance, prompting further investigation into the underlying causes.

One possible explanation is that user behavior and interest can fluctuate rapidly within a short time frame. In the test conversations, interaction patterns categorized by Koike et al.~\cite{koike2025drives} as Messy and Awkward were frequently observed.

In these cases, dialogue success or failure is often determined within a brief interaction window. Longer inputs capturing more extended interactions may introduce excessive variability in user behavior, complicating prediction.

The superior accuracy of short inputs suggests that user motivation at the start of an interaction plays a crucial role in predicting success~\cite{koike2025drives}. Early behavioral cues, such as a user approaching the robot directly, can indicate strong engagement and a higher likelihood of completing the interaction. These pre-interaction behaviors, along with early shifts in interest, are best captured through nonverbal cues, such as body orientation and group engagement. This further highlights the importance of high-accuracy VLM recognition, as discussed in the previous section.

\subsection{Evaluation under Different Robot Configurations}
To complement the results in Section~\ref{sec:results_in_offline_prediction_task}, we newly evaluated the performance under a different robot configuration. First, we collected a new route-guidance dialogue dataset. The dialogues were collected at a different installation location, and the robot’s behavior was changed by the prompts from the original behavior described in Section~\ref{sec:dataset_collenction}. The dataset comprises 20 short-input dialogues and 20 long-input dialogues, with 10 successful and 10 failed cases. Then, using the dialogue strategy repository acquired under the VLM-Both condition from the training data in Section~\ref{sec:experiment}, we conducted the offline prediction task on this new dataset.

As a result, the F1 score was 0.7 for the short-input and 0.85 for the long-input, demonstrating high performance. Although the performance for short and long inputs was reversed, these results suggest that the proposed architecture and the VLM-Both model exhibit consistently high accuracy even under different robot configurations.

\section{Conclusions}
\label{sec:conclusion}
This study introduced an architecture that integrates VLM and LLM to automatically acquire dialogue strategies by analyzing successful and failed interactions. Experimental results demonstrated that incorporating both success and failure strategies led to more accurate predictions of dialogue outcomes compared to approaches that relied solely on successful cases. These findings validate the proposed method's effectiveness in acquiring practical dialogue strategies without being constrained to a limited set of successful cases.

The results highlight that not all user attributes contribute equally to effective dialogue strategies: dynamic, task-relevant cues such as engagement were more informative than static identity-related attributes (e.g., age or gender). This underlines both the critical role of nonverbal information and the limited necessity—and potential ethical risk—of relying heavily on identity attributes. Future work will focus on integrating the acquired strategies into a dialogue robot that tracks transient user states in real time and adapts its behavior accordingly, while minimizing dependence on sensitive attributes. Such a design has potential applications beyond navigation tasks, including adaptive dialogue systems for customer service (adjusting responses based on user engagement or purchase intent) and medical consultations (optimizing strategies to address patient anxiety).

As a limitation of this study, the evaluation was restricted to the route guidance domain. Under open-ended conditions, such as small talk, dialogue strategies may have stronger influence; however, this influence may be beneficial or detrimental and may expose system limitations. Evaluating such scenarios would require richer interaction data from a new real-world deployment with a substantially different task design. In addition, GPT-4o was used as both the VLM and the LLM, and no performance comparison across different models was conducted. Therefore, the impact of model selection on the results remains to be investigated in future work.

\bibliographystyle{IEEEtran}
\bibliography{references}

@article{okafuji2022behavioral,
  author={Okafuji, Yuki and Ozaki, Yasunori and Baba, Jun and Nakanishi, Junya and Ogawa, Kohei and Yoshikawa, Yuichiro and Ishiguro, Hiroshi},
  journal={International journal of social robotics},
  number={7},
  pages={1731--1747},
  title={Behavioral assessment of a humanoid robot when attracting pedestrians in a mall},
  volume={14},
  year={2022},
}

@article{kass1988modeling,
  author={Kass, Robert and Finin, Tim},
  journal={Computational Linguistics},
  number={3},
  pages={5--22},
  title={Modeling the user in natural language systems},
  volume={14},
  year={1988},
}

@article{FONG2003143,
  author = {Terrence Fong and Illah Nourbakhsh and Kerstin Dautenhahn},
  journal = {Robotics and Autonomous Systems},
  number = {3},
  pages = {143--166},
  title = {A survey of socially interactive robots},
  volume = {42},
  year = {2003},
}

@article{janarthanam2014adaptive,
    author = {Janarthanam, Srinivasan and Lemon, Oliver},
    journal = {Computational Linguistics},
    number = {4},
    pages = {883--920},
    title = {Adaptive Generation in Dialogue Systems Using Dynamic User Modeling},
    volume = {40},
    year = {2014},
}

@article{VINCIARELLI20091743,
  author = {Alessandro Vinciarelli and Maja Pantic and Hervé Bourlard},
  journal = {Image and Vision Computing},
  number = {12},
  pages = {1743--1759},
  title = {Social signal processing: survey of an emerging domain},
  volume = {27},
  year = {2009},
}

@inproceedings{xie2024few,
  author={Xie, Zhouhang and Majumder, Bodhisattwa Prasad and Zhao, Mengjie and Maeda, Yoshinori and Yamada, Keiichi and Wakaki, Hiromi and McAuley, Julian},
  booktitle={Findings of ACL},
  title={Few-shot Dialogue Strategy Learning for Motivational Interviewing via Inductive Reasoning},
  year={2024},
  pages={13207--13219},
}

@article{kanda2010communication,
  author={Kanda, Takayuki and Shiomi, Masahiro and Miyashita, Zenta and Ishiguro, Hiroshi and Hagita, Norihiro},
  journal={IEEE Transactions on Robotics},
  number={5},
  pages={897--913},
  title={A communication robot in a shopping mall},
  volume={26},
  year={2010},
}

@inproceedings{yao2023react,
  author = {Yao, Shunyu and Zhao, Jeffrey and Yu, Dian and Du, Nan and Shafran, Izhak and Narasimhan, Karthik and Cao, Yuan},
  booktitle={Proc. ICLR},
  title = {{ReAct}: synergizing Reasoning and Acting in Language Models},
  year = {2023},
}

@inproceedings{liang2023code,
  author={Liang, Jacky and Huang, Wenlong and Xia, Fei and Xu, Peng and Hausman, Karol and Ichter, Brian and Florence, Pete and Zeng, Andy},
  booktitle={Proc. ICRA}, 
  pages={9493--9500},
  title={Code as Policies: language Model Programs for Embodied Control}, 
  year={2023},
}

@article{wang2023voyager,
  author={Guanzhi Wang and Yuqi Xie and Yunfan Jiang and Ajay Mandlekar and Chaowei Xiao and Yuke Zhu and Linxi Fan and Anima Anandkumar},
  journal={Transactions on Machine Learning Research},
  title={Voyager: an Open-Ended Embodied Agent with Large Language Models},
  year={2024},
}

@inproceedings{xu2024language,
  author = {Xu, Zelai and Yu, Chao and Fang, Fei and Wang, Yu and Wu, Yi},
  booktitle = {Proc. ICML},
  title = {Language agents with reinforcement learning for strategic play in the Werewolf game},
  year = {2024},
  pages = {55434--55464},
}

@article{su2016continuously,
  author={Su, Pei-Hao and Gasic, Milica and Mrksic, Nikola and Rojas-Barahona, Lina and Ultes, Stefan and Vandyke, David and Wen, Tsung-Hsien and Young, Steve},
  journal={arXiv preprint arXiv:1606.02689},
  title={Continuously learning neural dialogue management},
  year={2016},
}

@article{mctear2002spoken,
  author={McTear, Michael F},
  journal={ACM Computing Surveys},
  number={1},
  pages={90--169},
  title={Spoken dialogue technology: enabling the conversational user interface},
  volume={34},
  year={2002},
}

@article{paiva2017empathy,
  author={Paiva, Ana and Leite, Iolanda and Boukricha, Hana and Wachsmuth, Ipke},
  journal={ACM Transactions on Interactive Intelligent Systems},
  number={3},
  pages={1--40},
  title={Empathy in virtual agents and robots: a survey},
  volume={7},
  year={2017},
}

@inproceedings{devault2014simsensei,
  author = {DeVault, David and Artstein, Ron and Benn, Grace and Dey, Teresa and Fast, Ed and Gainer, Alesia and Georgila, Kallirroi and Gratch, Jon and Hartholt, Arno and Lhommet, Margaux and Lucas, Gale and Marsella, Stacy and Morbini, Fabrizio and Nazarian, Angela and Scherer, Stefan and Stratou, Giota and Suri, Apar and Traum, David and Wood, Rachel and Xu, Yuyu and Rizzo, Albert and Morency, Louis-Philippe},
  booktitle={Proc. AAMAS},
  pages={1061--1068},
  title={{SimSensei} Kiosk: a virtual human interviewer for healthcare decision support},
  year={2014},
}

@inproceedings{yamamoto2023character,
  author={Yamamoto, Kenta and Inoue, Koji and Kawahara, Tatsuya},
  booktitle={Proc. IWSDS},
  title={Character adaptation of spoken dialogue systems based on user personalities},
  year={2023},
}

@inproceedings{komatani2003flexible,
  author={Komatani, Kazunori and Ueno, Shinichi and Kawahara, Tatsuya and Okuno, Hiroshi G},
  booktitle={Proc. ACL},
  pages={256--263},
  title={Flexible guidance generation using user model in spoken dialogue systems},
  year={2003},
}

@article{del2022learning,
  author={Del Duchetto, Francesco and Hanheide, Marc},
  journal={IEEE Robotics and Automation Letters},
  number={3},
  pages={6934--6941},
  title={Learning on the job: Long-term behavioural adaptation in human-robot interactions},
  volume={7},
  year={2022},
}

@article{alayrac2022flamingo,
  author = {Alayrac, Jean-Baptiste and Donahue, Jeff and Luc, Pauline and Miech, Antoine and Barr, Iain and Hasson, Yana and Lenc, Karel and Mensch, Arthur and Millican, Katherine and Reynolds, Malcolm and Ring, Roman and Rutherford, Eliza and Cabi, Serkan and Han, Tengda and Gong, Zhitao and Samangooei, Sina and Monteiro, Marianne and Menick, Jacob L and Borgeaud, Sebastian and Brock, Andy and Nematzadeh, Aida and Sharifzadeh, Sahand and Bi\'{n}kowski, Miko\l aj and Barreira, Ricardo and Vinyals, Oriol and Zisserman, Andrew and Simonyan, Kar\'{e}n},
  journal={Advances in neural information processing systems},
  pages={23716--23736},
  title={Flamingo: a visual language model for few-shot learning},
  year={2022},
}

@inproceedings{li2023blip2,
  author={Li, Junnan and Li, Dongxu and Xiong, Caiming and Hoi, Steven C.H.},
  booktitle={Proc. ICML},
  title={{BLIP-2}: bootstrapping Language-Image Pre-training for Unified Vision-Language Understanding and Generation},
  year={2023},
  pages={19730--19742},
}

@inproceedings{koike2025drives,
  author={Koike, Amy and Okafuji, Yuki and Hoshimure, Kenya and Baba, Jun},
  booktitle={Proc. HRI},
  title={What Drives You to Interact?: the Role of User Motivation for a Robot in the Wild},
  year={2025},
  pages={183--192},
}

@article{walker2000application,
  author={Walker, Marilyn A},
  journal={Journal of Artificial Intelligence Research},
  pages={387--416},
  title={An application of reinforcement learning to dialogue strategy selection in a spoken dialogue system for email},
  volume={12},
  year={2000},
}

@inproceedings{scheffler2022automatic,
  author = {Scheffler, Konrad and Young, Steve},
  booktitle = {Proc. HLT},
  pages = {12--19},
  title = {Automatic learning of dialogue strategy using dialogue simulation and reinforcement learning},
  year = {2002},
}

@inproceedings{finch-choi-2020-towards,
  author={Finch, Sarah E. and Choi, Jinho D.},
  booktitle={Proc. SIGDIAL},
  pages={236--245},
  title={Towards Unified Dialogue System Evaluation: a Comprehensive Analysis of Current Evaluation Protocols},
  year={2020},
}

@book{kendon1990conducting,
  author={Kendon, Adam},
  publisher={CUP Archive},
  title={Conducting interaction: patterns of behavior in focused encounters},
  volume={7},
  year={1990},
}

@article{sakaguchi2022estimation,
  author={Sakaguchi, Taichi and Okafuji, Yuki and Matsumura, Kohei and Baba, Jun and Nakanishi, Junya},
  journal={arXiv preprint arXiv:2206.02394},
  title={An estimation framework for passerby engagement interacting with social robots},
  year={2022},
}

@inproceedings{lu2019learnfailure,
    author = {Lu, Keting and Zhang, Shiqi and Chen, Xiaoping},
    title = {Goal-oriented dialogue policy learning from failures},
    year = {2019},
    booktitle = {Proc. AAAI},
    volume = {33},
    number={01},
    pages={2596--2603},
}

@inproceedings{delduchetto2023fail,
  author    = {Del Duchetto, Francesco and Kucukyilmaz, Alper and Hanheide, Marc},
  title     = {In-the-Wild Failures in a Long-Term {HRI} Deployment},
  booktitle = {Proc. ICRA},
  year      = {2023},
  note      = {Poster},
}

@inproceedings{williams2023eye,
    author = {Williams, Tom},
    title = {The Eye of the Robot Beholder: ethical Risks of Representation, Recognition, and Reasoning over Identity Characteristics in Human-Robot Interaction},
    year = {2023},
    booktitle = {Companion of HRI},
    pages = {1--10},
}

@article{Foggia2024SoftBiometrics,
  author  = {Foggia, Pasquale and Greco, Antonio and Roberto, Antonio and Saggese, Alessia and Vento, Mario},
  title   = {Identity, Gender, Age, and Emotion Recognition from Speaker Voice with Multi-task Deep Networks for Cognitive Robotics},
  journal = {Cognitive Computation},
  volume  = {16},
  pages   = {2713--2723},
  year    = {2024},
}

\end{document}